\documentclass[aps,prb,twocolumn,groupedaddress]{revtex4-2}

\usepackage{amssymb}
\usepackage{amsmath}
\usepackage{bm}
\usepackage{graphicx}
\usepackage{dcolumn}
\usepackage{longtable}
\usepackage{color}
\usepackage{multirow}

\newcommand{\rsp}{black}
\newcommand{\PbC}{Pb(TiO)Cu$_4$(PO$_4$)$_4$}
\newcommand{\BaC}{Ba(TiO)Cu$_4$(PO$_4$)$_4$}
\newcommand{\SrC}{Sr(TiO)Cu$_4$(PO$_4$)$_4$}

\begin{document}


\title{High-field NMR study of field-induced states in \PbC}

\author{Y.~Ihara}
\email{yihara@phys.sci.hokudai.ac.jp}
\affiliation{Department of Physics, Faculty of Science, Hokkaido University, Sapporo 060-0810, Japan}
\author{T.~Kanda}
\altaffiliation{Current address: Department of Basic Science, Graduate School of Arts and Sciences, The University of Tokyo, Tokyo 153-8902, Japan}
\affiliation{Institute for Solid State Physics, The University of Tokyo, Chiba, 277-8581, Japan}
\author{Y.~Kato}
\affiliation{Department of Applied Physics, University of Fukui, Fukui 910-8507, Japan}
\author{Y.~Motome}
\affiliation{Department of Applied Physics, The University of Tokyo, Tokyo, 113-8656, Japan}
\author{K.~Matsui}
\affiliation{Institute for Solid State Physics, The University of Tokyo, Chiba, 277-8581, Japan}
\author{K.~Kindo}
\affiliation{Institute for Solid State Physics, The University of Tokyo, Chiba, 277-8581, Japan}
\author{Y.~Kohama}
\affiliation{Institute for Solid State Physics, The University of Tokyo, Chiba, 277-8581, Japan}
\author{T.~Kimura}
\affiliation{Department of Applied Physics, The University of Tokyo, Tokyo, 113-8656, Japan}
\author{K.~Kimura}
\affiliation{Department of Materials Science, Osaka Metoropolitan University, Osaka, 599-8531, Japan}
\date{\today}

\begin{abstract}
The square cupola antiferromagnet \PbC\, exhibits the intriguing magnetoelectric responses arising from the consecutive change in the magnetic quadrupolar-type configuration of magnetic moments under external magnetic fields higher than 15 T. 
To clarify the high-field magnetic structures in \PbC, an NMR measurement was performed in pulsed fields up to 32.2~T significantly extending the field range accessible by superconducting magnets.
The double-peak structure of NMR spectra emerging above 29 T applied along the [001] direction evidences the successive magnetic transitions.  
The field dependence of NMR spectra was analyzed on the basis of cluster mean-field theory, which allows us to propose possible magnetic structures for the high-field magnetic states. 
\end{abstract} 

\maketitle

\section{Introduction}

Electrons in crystal lattice obtain composite degrees of freedom from the combination of their spin, charge and orbital degrees of freedom.
The higher-order degrees of freedoms, which are described by the multipole moments \cite{Kuramoto-2008}, exhibit interesting electric and magnetic responses demonstrating the potential to convert  information between magnetic and  electric devices \cite{Spaldin-NatMat18}. 
The extended multipole is carried by a cluster of several spins, on which a particular arrangement of spins is introduced by the competing interactions, such as exchange, spin-orbit, and Dzyaloshinsky-Moriya interactions \cite{Spaldin-JPCM20, Spaldin-PRB88, Suzuki-PRB99}. 
The clarification of their response to magnetic fields is essential for the progress toward the practical application.

\begin{figure}[tbp]
\begin{center}
\includegraphics[width=7.0cm]{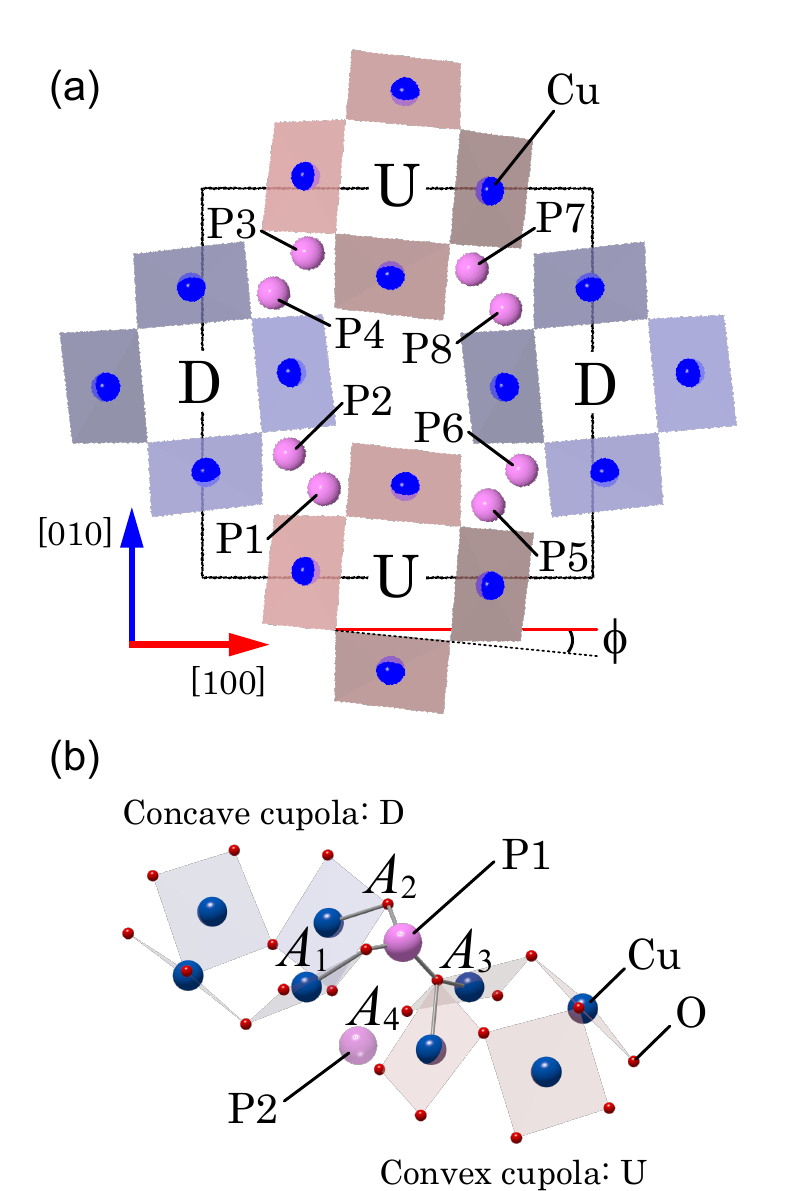}
\end{center}
\caption{(a) Crystal structure of \PbC\, viewed from the [001] axis.
Convex and concave Cu$_4$O$_{12}$ clusters are labeled as U and D, respectively. 
Oxygen sites are located at each corner of rectangles but omitted for clarity. 
The thin black line represents the unit cell. 
Eight P sites in the unit cell are shown by pink balls and labeled as P1 to P8. 
$\phi$ is the rotation angle of cupola cluster from the [100] axis.  
(b) A closer view around P1 and P2 sites.
One P site is surrounded by four neighboring Cu sites. 
The hyperfine coupling constants between these Cu$^{2+}$ magnetic moments and P1 nuclear moment are labeled as $A_1$ to $A_4$ following the order of distance (see text for detail).  }
\label{structure}
\end{figure}

The magnetoelectric (ME) response of the magnetic cluster was initially unveiled on Cr$_2$O$_3$ with antiparallel spin pairs, which are described by the sum of magnetic monopole and quadrupole moments \cite{Folen-PRL1961}.
Subsequent studies revealed the ME response originating from combined magnetic quadrupole and toroidal moments in GaFeO$_3$ \cite{Arima-PRB2004} and further extended to those induced by the vector spin chirality of incommensurate magnetic structures in TbMnO$_3$ \cite{Kimura-Nature2003}, but the purely quadrupolar-type spin configuration had remained elusive. 
The magnetic quadrupolar moment was discovered in a multiferroic material $A$(TiO)Cu$_4$(PO$_4$)$_4$ ($A=$ Ba, Sr, and Pb) \cite{Kimura-NatComm2016, Kimura-PRB2018}, in which a specific spin configuration is realized on the square cupola structure of Cu$_4$O$_{12}$ clusters. 
A dome-like structure of Cu$_4$O$_{12}$ clusters is constructed by four walls of CuO$_4$ plaquettes (Fig.~\ref{structure}). 
When magnetic moments of Cu$^{2+}$ ions on each wall order antiferromagnetically by the dominant nearest neighbor interaction, the two-in-two-out spin configuration is realized due to the steric angles between CuO$_4$ plaquettes and the spin-orbit coupling that fixes the spin direction perpendicular to the CuO$_4$ plane.

The antiferromganetic (AFM) structure at zero magnetic field was solved by previous neutron and NMR measurements \cite{Kimura-NatComm2016,Babkevich-PRB96, Islam-PRB2018,Kumar-JMMM492,Rasta-PRB2020}. 
The ME response associated with the magnetic quadrupole moment was observed in Pb(TiO)Cu$_4$(PO$_4$)$_4$ \cite{Kimura-PRB2018}. 
Namely, the magnetic-field induced electric polarization appears along the $[110]$ direction while the magnetic field is applied to the $[1 \bar{1} 0]$ direction characterizing the off-diagonal components of ME tensor for the magnetic quadrupole moment \cite{Hayami-PRB98}.  
The magnetic structure at low magnetic fields is modified by applying extremely high magnetic fields and the ME activity is modified accordingly \cite{Kimura-JPSJ2019,Kimura-PRM2018}. 
The evolution of magnetic and electric properties were consistently explained by a cluster mean-field theory that incorporates several realistic interactions \cite{Kato-PRL2017,Kimura-PRM2018, Kato-PRB2019}.
The magnetic isotherm and the field variation of electric polarization calculated for $A$(TiO)Cu$_4$(PO$_4$)$_4$ fit well to the experimental results. 
The success in the theoretical calculation leads to a realistic spin model for the field-induced states. 
An exception is, however, the peak in the dielectric constant at 26 T for fields parallel to the [001] direction \cite{Miyake-RSI2020}.
The apparent peak in the dielectric constant clearly points out a field-induced phase transition, but no such phase transition was predicted from the theory. 
The magnetization shows a broad hump around the corresponding field range, suggesting that the magnetic moment along the [001] direction is merely affected at the transition in contrast to the significant modification of electric polarization \cite{Kimura-PRM2018}.
Practically, the selective control only for electric polarization will allow us to invent a methodology useful for writing electric memory independent from magnetic memory in a multiferroic device. 
The experimental investigation of the field-induced magnetic structure allows us to understand the field response of the multiferroic material and to theoretically explain the field-induced phase transitions more consistently. 
In \PbC, however, as the metamagnetic transitions occur in high field region above 15 T, experimental observation of the microscopic magnetic structure has never been achieved. 

In this study, we have performed NMR spectroscopy under pulsed high magnetic fields up to 32.2 T and investigated the magnetic structures of the field-induced phases in \PbC\, by simulating the internal fields on the basis of cluster mean-field theory. 
The NMR experiment is crucial to microscopically study the magnetic properties in high magnetic fields.
The pulsed field expands the accessible field range up to several tens of tesla, which is hard to generate using a superconducting magnet.  
Although the temporal change in the field strength imposes technical difficulty to observe NMR signal, the flat-top pulsed field extensively facilitates the NMR spectrum measurement in magnetic fields even greater than 50 T \cite{Ihara-RSI2021,Kohama-JAP2022, Kuhne-ContPhys2024}.
The NMR spectra obtained under the feedback-controlled flat-top pulsed fields clearly detect the modification of magnetic structures associated with the metamagnetic transitions at high fields.  
We also performed the NMR measurement at low magnetic fields to confirm the magnetic quadrupolar-type spin structure previously suggested by the neutron diffraction studies \cite{Kimura-NatComm2016}.
Our NMR study provides information complimentary to the neutron diffraction study, namely the local arrangement of spins such as quadrupolar moments on the Cu$_4$O$_{12}$ clusters.  

This manuscript is organized as follows.
After the introduction, the detailed experimental procedures including \textcolor{\rsp}{pulsed}-field NMR measurement are described in \S~\ref{setup}. 
In \S~\ref{res}, the NMR results in the paramagnetic state, low-field ordered state and high-field ordered state are presented. 
Then, the magnetic structure in the high-field phase is discussed in \S~\ref{dis} based on the field dependence of NMR spectra.
We will propose a magnetic structure for unsolved high-field phase supported by the cluster mean-field theory. 
Finally, this study is summarized in \S~\ref{con}.

\section{Experimental}\label{setup}
The crystal structure of \PbC\, has chirality associated with the rotation of Cu$_4$O$_{12}$ cluster \cite{Kimura-InorgChem2016}.
The rotation angle $\phi$ of an upward cupola cluster is indicated in Fig.~\ref{structure}(a). 
The single crystalline sample measured in this study was grown by the slow cooling method and cut into a square shape with a dimension of $4\times 4\times 1$ mm$^3$. 
To check the chirality of the sample, a sense of optical rotation was examined using polarized light microscopy in transmission geometry. 
The whole area of the sample showed dextrorotation, confirming that the sample is monochiral, with the upward cupolas rotating clockwise as shown in Fig.~\ref{structure}
\cite{Kimura-InorgChem2016}.

The $^{31}$P-NMR spectrum measurement at low fields was performed both in the frequency-sweep and field-sweep modes. 
At high temperatures above 90 K, where the NMR spectrum width is narrower than the frequency bandwidth of the NMR spectrometer, the NMR frequency spectra were obtained at a fixed field of approximately 13 T.
The spin echo signal was Fourier transformed (FT) to obtain the FT spectra against the frequency axis.  
At lower temperatures with broader spectral width, several transient FT spectra were combined to construct the whole spectrum. 
We also measured the field-sweep NMR spectra near and below $T_N$, for which the transient FT spectra at a fixed frequency were obtained during field sweep and their intensity was mapped on the field axis.
\textcolor{\rsp}{The field strength at each moment of FT spectra measurement was derived from the current flowing through the superconducting coil.}

The field-induced magnetic structures above 15 T were explored by the NMR spectrum measurements in pulsed magnetic fields. 
The frequency-sweep NMR spectrum was obtained in the flat-top pulsed fields equipped with the feedback control circuit, which allows us to suppress the temporal field variation and to reproducibly generate the same magnetic field strength \cite{Kohama-RSI2015}.
\textcolor{\rsp}{The field strength was measured by the $^{31}$P-NMR frequency of the sample at 77 K, where the narrow spectral width and known $K$ enable the precise field calibration.}
To measure the frequency-sweep NMR spectrum with rather broad total spectral width, the transient FT spectra were obtained at the same magnetic field but different center frequencies and recombined on the frequency axis.
Since the FT spectra should be measured at a constant external fields, the field strength of pulsed field must be kept constant during the measurement, which was challenging for conventional pulsed fields.   
The time duration for the quasi-steady field condition is typically 3 ms at the highest field, while it increases at lower fields, where the field margin for feedback control is large. 
Up to 10 transient FT spectra were obtained within one field pulse, taking advantage of the fast frequency skipping feature of the software defined radio based NMR spectrometer \cite{Ihara-RSI2021}.  
To minimize the overlap of measurement frequency we used the shaped radio-frequency pulse, which specifically irradiates the frequency range determined by the bandwidth of typically 200 kHz.

\section{Results} \label{res}

\begin{figure}[tbp]
\begin{center}
\includegraphics[width=7.0cm]{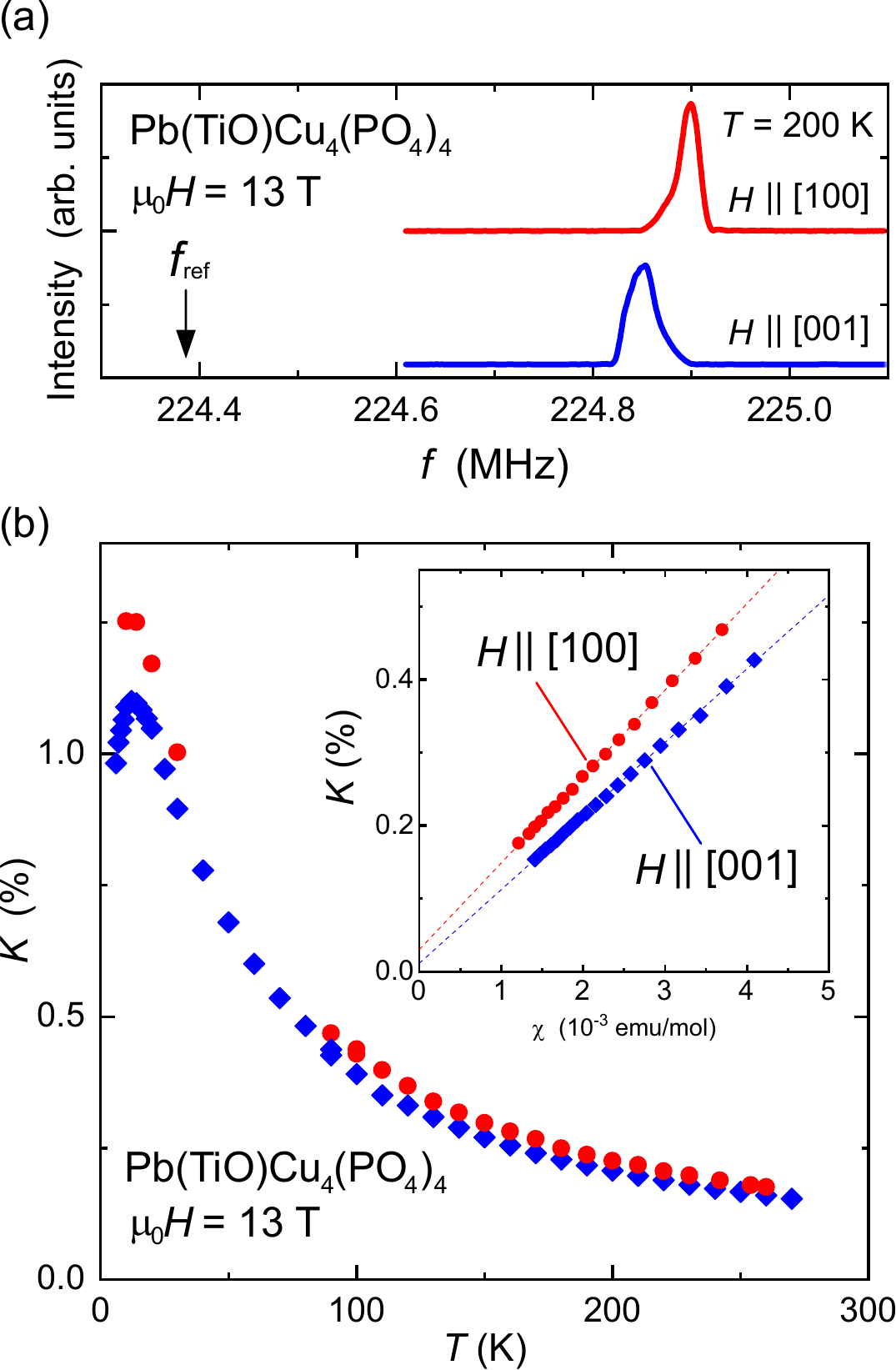}
\end{center}
\caption{(a)  $^{31}$P-NMR spectra obtained at 200 K in fields applied along the [100] (top, red) and [001] (bottom, blue) directions.
The reference frequency $f_{\rm ref}$ calculated from the field strength and $\gamma$ is indicated by downward arrow. 
(b) Temperature dependence of Knight shift for the [100] (circles, red) and [001] (squares, blue) directions.
Inset shows the $K-\chi$ plot, for which the $K(T)$ is plotted against $\chi(T)$ using the temperature as an implicit parameter. 
The hyperfine coupling constants are obtained from the linear relation. }
\label{kchi}
\end{figure}
\subsection{Paramagnetic state}

In this study, the $^{31}$P nuclear moment (nuclear spin $I=1/2$, gyromagnetic ratio $\gamma=$ 17.237 MHz/T) was used as a microscopic magnetic probe to investigate the magnetic structure in the AFM state based on the NMR measurement.  
The coupling constants between the electronic magnetic moments and nuclear moments were quantitatively estimated in the paramagnetic state to measure the internal fields generated by the ordered Cu$^{2+}$ moments in the AFM state. 
Crystallographically, \PbC\, has only one P site at [0.7252, 0.1847, 0.7441]. 
The symmetry operations in $P42_12$ space group generate eight P sites in a unit cell as shown and labeled in Fig.~\ref{structure}(a). 
Figure~\ref{structure}(b) illustrates the local configuration around the P1 site located between two cupola clusters directing upward (convex cupola: U) and downward (concave cupola: D).
Each P site is coupled to four neighboring Cu moments through the Cu-O-P exchange paths.  
The coupling strengths to each Cu moment are different, reflecting the site-dependent distances between P and Cu ranging from 3.039 \AA\; to 3.153 \AA.
The coupling constants are, then, labeled as $A_1$, $A_2$, $A_3$, and $A_4$ following the order of P-Cu distances. 

In the paramagnetic state, Cu$^{2+}$ moments are uniformly aligned to the external field direction creating a local magnetization $M_p$ at each Cu site.
The hyperfine field from the paramagnetic moments is written as $B_p = (A_1+A_2+A_3+A_4)M_p = A_pM_p$ using the total coupling constant $A_p$. 
By dividing the above equation by the applied magnetic field, the relationship between the Knight shift $K$ and the susceptibility $\chi$ is derived as $K(T) = A_p\chi(T)$. 
Experimentally, $A_p$ is estimated from the temperature dependence of $\chi$ and $K$.  
$K$ shows an anisotropy as demonstrated in Fig.~\ref{kchi}(a) by the difference in peak positions between the $^{31}$P-NMR spectra in $H\parallel[100]$ and $H\parallel[001]$ at high temperature.   
Figure \ref{kchi}(b) shows $K(T)$ measured in a fixed field of 13 T applied along the [100] and [001] directions.
To verify the linear relationship between $K(T)$ and $\chi(T)$, $K$ is plotted against $\chi$ measured at the corresponding temperatures \textcolor{\rsp}{above 90 K} in the inset of Fig.~\ref{kchi}(b). 
\textcolor{\rsp}{Since $\chi$ was measured at 5 T for both directions, we used high temperature data to eliminate the field dependence of $K$ and $\chi$.}
From the slope of the linear relation, the coupling constants for the [100] and [001] directions are estimated to be 662 mT/$\mu_B$ and 564 mT/$\mu_B$, respectively. 
The experimentally obtained coupling constants include a dipole coupling in addition to the hyperfine coupling. 
The dipole contribution is calculated from the structure, and the result yields $-12$ mT/$\mu_B$ and $7$ mT/$\mu_B$ for the [100] and [001] directions, respectively. 
These contributions are subtracted from the total coupling constants. 
$A_p$ for the [100] and [001] directions are, thus, $674$ mT/$\mu_B$ and $557$ mT/$\mu_B$. 
The isotropic coupling constant of $A_p = 635$ mT/$\mu_B$ is obtained from these values and is comparable to those reported for sister compounds \cite{Islam-PRB2018,Kumar-JMMM492,Rasta-PRB2020}.  
For the further analyses in the AFM state, we use the dominant isotropic term neglecting a smaller anisotropic term, which originates from a small population of $p$ electrons at P site.

\begin{figure}[tbp]
\begin{center}
\includegraphics[width=6.5cm]{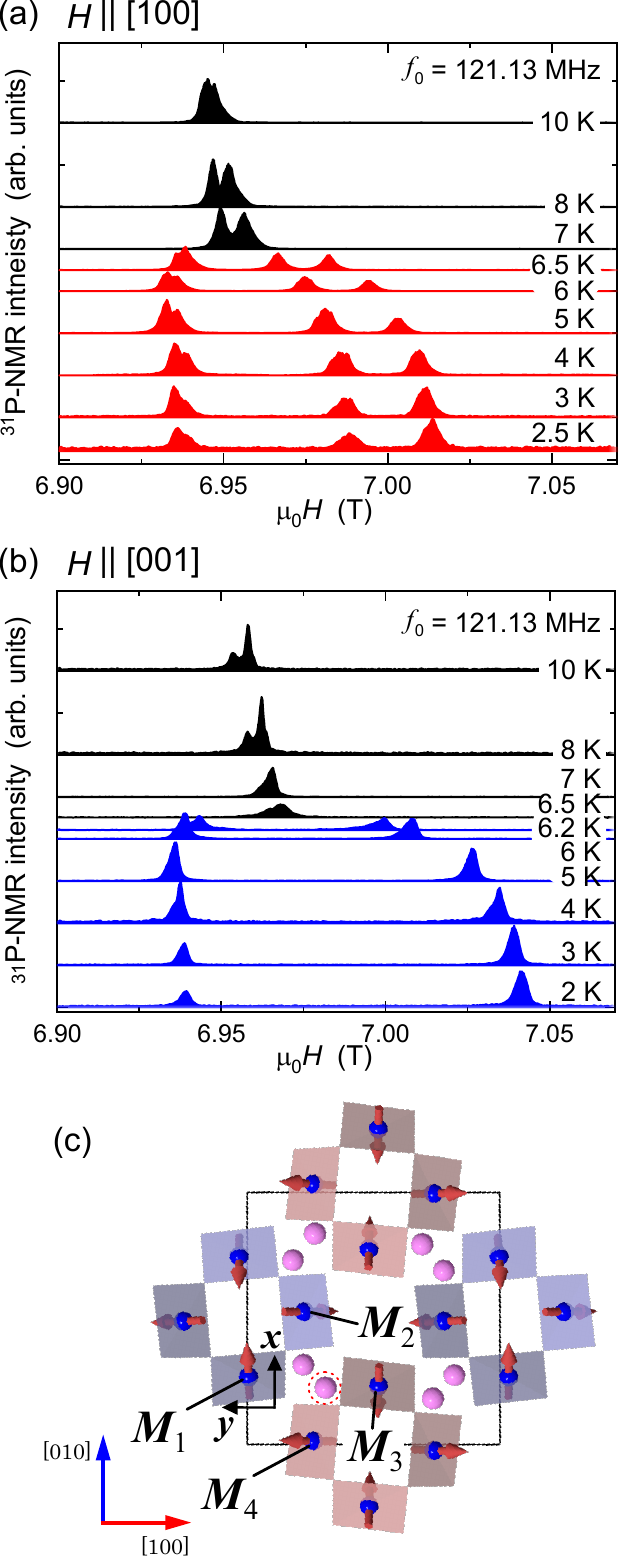}
\end{center}
\caption{Field-sweep NMR spectra around $T_N$ for fields along the (a) [100] and (b) [001] directions.
Spectrum splitting observed below $T_N$ is caused by the spontaneous internal fields in the low-field AFM state (LF phase).
(c) Magnetic structure of the quadrupolar type AFM state. 
The four Cu$^{2+}$ moments neighboring to the P site marked by dotted circle are labeled as $M_1 \sim M_4$. 
The local $xy$ axes are defined around $M_1$. }
\label{NMRSP_LF}
\end{figure}

\subsection{Low-field antiferromagnetic structure}
From the analyses in the paramagnetic state, $A_p$ cannot be decomposed into the site-dependent $A_i (i=1\sim 4)$, which are required to investigate the magnetic structures in the field-induced magnetic states.  
We extend our analyses to the AFM state at low magnetic fields (LF phase), in which four magnetic moments in a cupola construct the magnetic quadrupole moment \cite{Kimura-NatComm2016}. 
The non-colinear and non-coplanar spin configurations allow us to access the site-dependent coupling constants.  
The $^{31}$P-NMR spectra in the LF phase were measured around 7 T, which is sufficiently small to maintain the magnetic quadrupolar-type magnetic structure \cite{Kimura-JPSJ2019}.
We chose the field-sweep method with a constant frequency of 121.13 MHz to measure the $^{31}$P-NMR spectra at low fields.
The field-sweep method can flexibly cover the field range required to capture the full spectra split by the internal fields. 
Figures \ref{NMRSP_LF} (a) and (b) show the temperature variation of NMR spectra below 10 K for the [100] and [001] directions, respectively. 
The clear peak splitting was observed in the LF phase below $T_N\simeq 7$ K, confirming that the $^{31}$P nuclear moments are affected by the staggered fields generated in the LF phase. 
The single peak in the paramagnetic state splits into three peaks for $H \parallel [100]$ and two for $H \parallel [001]$.
The three peaks for $H \parallel [100]$ are interpreted as the superimposed two sets of two peaks as observed in the case of \BaC \cite{Rasta-PRB2020} and as explained below in detail. 
The small peak splitting observed above $T_N$ for $H \parallel [100]$ is attributed to the development of short-range correlations and associated formation of local AFM spin structures, which locally differentiate the [100] and [010] directions for P sites connected by 4-fold symmetry, {\it e.g.} P1 and P4 sites. 
A small peak near the main peak for $H \parallel [001]$ comes from a small (5 degrees at most) misalignment of the crystalline axis with respect to the rotation axis of our single-axis rotator. 

From the peak separations observed at the lowest temperature, the internal field $\bm{B}_{\rm int}$ at P sites were measured for all three directions to be ${\bm B}_{\rm int}  = (38, 27, 51)$ mT. 
The internal field components along the [100] and [010] directions were simultaneously measured by the $^{31}$P-NMR spectra in $H\parallel [100]$, because $H \parallel [100]$ for the P1 sites is equivalent to $H \parallel [010]$ for the P4 sites. 
The contribution of the dipole coupling was calculated by summing up the direct dipole fields from the Cu moments within 100 \AA\, from the target P sites.
The resulting dipole field of $\bm{B}_{\rm dip} = (24,36,3.8)$ mT was subtracted from the experimentally obtained $\bm{B}_{\rm int}$ to estimate the hyperfine coupling contributions to be $\bm{B}_{\rm hf} = (B_x, B_y, B_z)= (13, -10, 48)$ mT. 
The hyperfine field is generated from four neighboring Cu$^{2+}$ moments, which construct the magnetic quadrupolar-type structure. 
The directions of magnetic moments are identified by the symmetry operation in the magnetic space group $P4'2_12'$ suggested by the previous neutron diffraction study \cite{Kimura-NatComm2016}.
To explicitly formulate the hyperfine field at P1 site using the hyperfine coupling constants, we defined the magnetic moment $\bm{M}_1$ in Fig.~\ref{NMRSP_LF}(c) as $\bm{M}_1 = (M_x, M_y, M_z)$. 
Here, $x$, $y$, and $z$ are parallel to the $[010], [ \bar{1} 00]$, and [001] directions, respectively. 
The magnetic moments at the other Cu sites are written with these components.
Then, the hyperfine field at P1 site is derived as  
\begin{align}
B_x &= (A_1-A_3)M_x - (A_2+A_4)M_y, \notag \\ 
B_y &= (A_2-A_4)M_x+(A_1+A_3)M_y, \notag \\ 
B_z &= (A_1-A_2-A_3+A_4)M_z. \notag 
\end{align}
From these equations and the total hyperfine coupling constant $A_p$, $M_y$ is written as 
\begin{align}
M_y = \frac{1}{A_p}\left[ (B_y-B_x)+B_z\frac{M_x}{M_z}\right]. \label{My} 
\end{align}
Since $A_p$ is one order of magnitude larger than each component of $\bm{B}_{\rm hf}$, eq.~(\ref{My}) suggests a small value for $M_y$, which is in line with the magnetic quadrupolar spin configuration and excludes the magnetic structure with large $M_y$, proposed by the previous neutron diffraction measurement as a possible structure \cite{Kimura-NatComm2016}.
For the following analyses, we assume ${M}_x = 0.48\;\mu_B$ and ${M}_z = 0.64\;\mu_B$ as estimated by the neutron diffraction measurement, but neglect the small $M_y$ for simplicity.  

The total hyperfine coupling constant $A_p= 635$ mT/$\mu_B$ is decomposed to site-dependent coupling constants $A_i$ through the relationship between $A_i$ and $\bm{B}_{\rm hf}$. 
The shorter bonding path between P nuclear moment and Cu$^{2+}$ moment is expected to host the stronger coupling, thus $A_i>A_{i+1}$. 
On the other hand, a negative value of $B_y$ suggests $A_2<A_4$. 
We also assume $A_3=A_4$ because the coupling paths to $\bm{M}_3$ ($A_3$) and $\bm{M}_4$ ($A_4$) share the same P-O bond [Fig.~\ref{structure}(b)]. 
This consideration results in $A_1>A_3=A_4>A_2$
and gives the following estimate of $A_i$, 
\begin{align}
A_1 &=185\; {\rm mT}/\mu_B, \notag \\
A_2 &=136\; {\rm mT}/\mu_B, \notag \\
A_3 &= A_4 = 157\; {\rm mT}/\mu_B. \notag 
\end{align}
These values are used for the analyses of magnetic structure in high magnetic fields.

\subsection{Field-induced magnetic phases}
\begin{figure}[tbp]
\begin{center}
\includegraphics[width=\columnwidth]{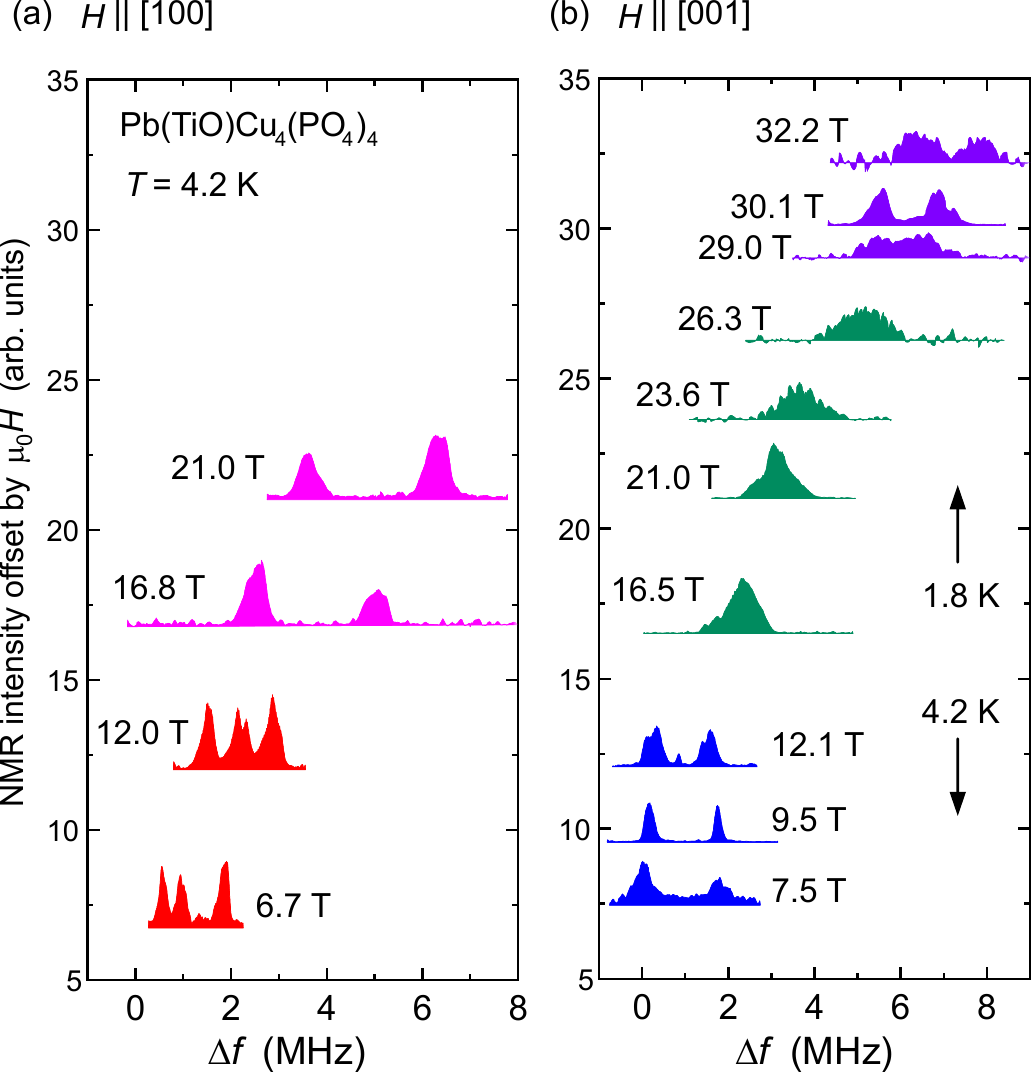}
\end{center}
\caption{Field dependence of $^{31}$P-NMR spectra at low temperatures measured in fields along the (a) [100] and (b) [001] directions. 
The frequency-sweep spectra were obtained in the pulsed field with the flattop option. 
For $H\parallel[100]$, two peaks were observed \textcolor{\rsp}{ at high fields characterizing phase Y \cite{Kato-PRL2017}. 
The critical field coincides with that determined from the dielectric constant measurement  $B_{a}= 14.8$ T \cite{Kimura-JPSJ2019}.}
For $H\parallel[001]$, the measurement temperature is reduced to 1.8 K for fields higher than 13 T. 
The successive modifications of spectral shape were observed \textcolor{\rsp}{above 12.1 T and 29.0 T. 
The lower field is consistent with the critical field determined by dielectric constant measurement, $B_{c1}=12.3$ T \cite{Kimura-JPSJ2019}. 
The high-field transition is attributed to the  unsolved dielectric anomaly detected at 26 T  \cite{Miyake-RSI2020}.}} 
\label{NMRSP_HF}
\end{figure}

The field-induced states at high magnetic fields are investigated by the NMR measurement in the pulsed high fields up to 32.2 T. 
The frequency-sweep NMR spectra obtained in the pulsed fields are summarized in Fig.~\ref{NMRSP_HF}. 
The horizontal axis of Fig.~\ref{NMRSP_HF} is the frequency shift $\Delta f$ from the reference frequency $f_0$ at the applied field strength $\mu_0H_0$ ($f_0 = \gamma \mu_0H_0$).
$\Delta f$ stems from the internal fields at each P site generated through the hyperfine and dipole couplings to the ordered Cu$^{2+}$ moments. 
The NMR spectra were measured in the AFM state at 4.2 K for $H\parallel [100]$.
The measurement temperature was further reduced to 1.8 K for $H \parallel [001]$ when the field is higher than 13 T, as the AFM transition temperature is suppressed to below 4 K \cite{Kimura-JPSJ2019}.
Previous study for magnetocaloric effect in \SrC\, revealed that the sample temperature decreases during the field pulse \cite{Nomura-PRB2023}.
This result confirms that the NMR spectra shown in Fig.~\ref{NMRSP_HF} were measured in the AFM ordered state, 
although the sample temperature was not monitored during the field pulse.

When the external field is applied along the $[100]$ direction [Fig.~\ref{NMRSP_HF}(a)], the three-peak spectrum was observed at the lowest field of 6.7 T in good agreement with the result obtained in the steady field presented in Fig.~\ref{NMRSP_LF}(a). 
Note that in the frequency-sweep spectrum, a positive internal field shifts a peak to the higher frequency, while that causes a negative shift in the field-sweep spectrum. 
Therefore, overlapped peaks appear at higher frequency in Fig.~\ref{NMRSP_HF}(a), while those are at lower fields in Fig.~\ref{NMRSP_LF}(a). 
The peak positions smoothly shift to the higher frequencies at higher fields until the critical field of $B_a = 14.8$ T \cite{Kimura-JPSJ2019}.
In the field-induced state above $B_a$, the clearly split two peaks were observed. 
Abrupt change in the spectral shape evidences the modification of magnetic structure associated with the metamagnetic transition. 
The internal fields in the LF and field-induced phases will be quantitatively analyzed in the next section. 

The NMR spectra in fields along the $[001]$ direction are shown in Fig.~\ref{NMRSP_HF}(b).
The double-peak spectra in the LF phase is consistent with the field-sweep spectra in steady fields [Fig.~\ref{NMRSP_LF}(b)]. 
An additional small splitting originates from a slight misalignment of the field direction from [001]. 
In the intermediate fields between $B_{c1}=12.3$ T \cite{Kimura-JPSJ2019} and $26.3$ T, the peak separation is reduced and the broad spectra without clearly resolved peaks were observed. 
The modification in the spectral shape above $B_{c1}$ is ascribed to the metamagnetic transition already detected by the dielectric constant and magnetization measurements \cite{Kimura-JPSJ2019}.
Remarkably, we found another abrupt change in the NMR spectral shape at fields higher than $B_{c2} = 29.0$ T, which corresponds to an unsolved dielectric anomaly \cite{Miyake-RSI2020, Nomura-PRB2023}.
The single broad spectrum splits into two peaks again above $B_{c2}$ evidencing a change in the magnetic structure, even \textcolor{\rsp}{though} the bulk magnetization shows only a broad hump around $B_{c2}$ \cite{Kimura-PRM2018}.  
In the next section, we will address the magnetic structure in the field-induced phases by quantitatively simulating the internal fields based on the magnetic structure model suggested by the theoretical study \cite{Kato-PRL2017, Kato-PRB2019}.

\section{Discussion} \label{dis}

\begin{figure}[tbp]
\begin{center}
\includegraphics[width=\columnwidth]{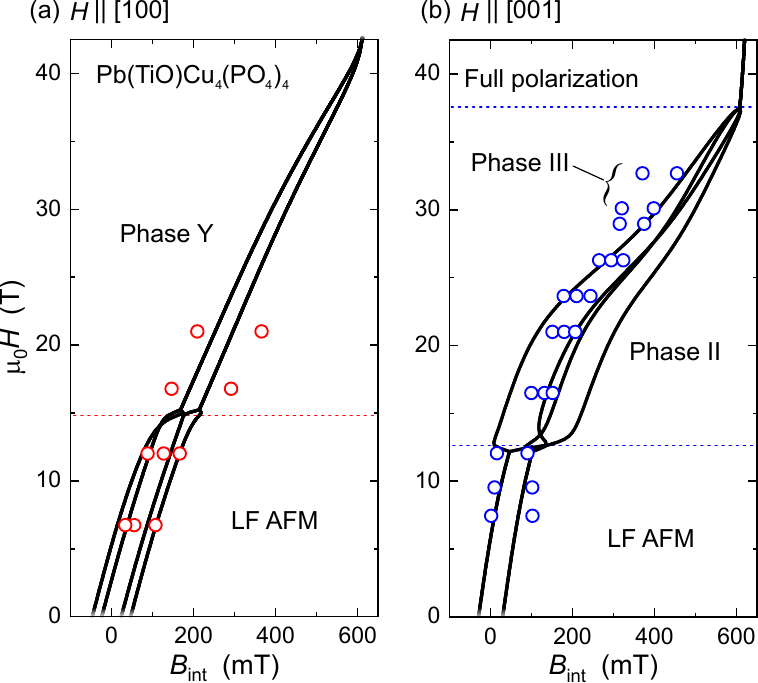}
\end{center}
\caption{
Internal field at P site calculated for the magnetic structures suggested by the CMF theory. 
The NMR peak positions are plotted by circle symbols. 
(a) When the field direction is [100], four (two) peaks are found in LF (Y) phase in line with experimental observation. 
(b) In fields along the [001] direction, NMR peak positions in LF AFM and phase II are consistently explained, while those above 29 T deviate from the theoretical expectation, suggesting the appearance of phase III.  
}
\label{Bint_cal}
\end{figure}

\begin{figure}[tbp]
\begin{center}
\includegraphics[width=\columnwidth]{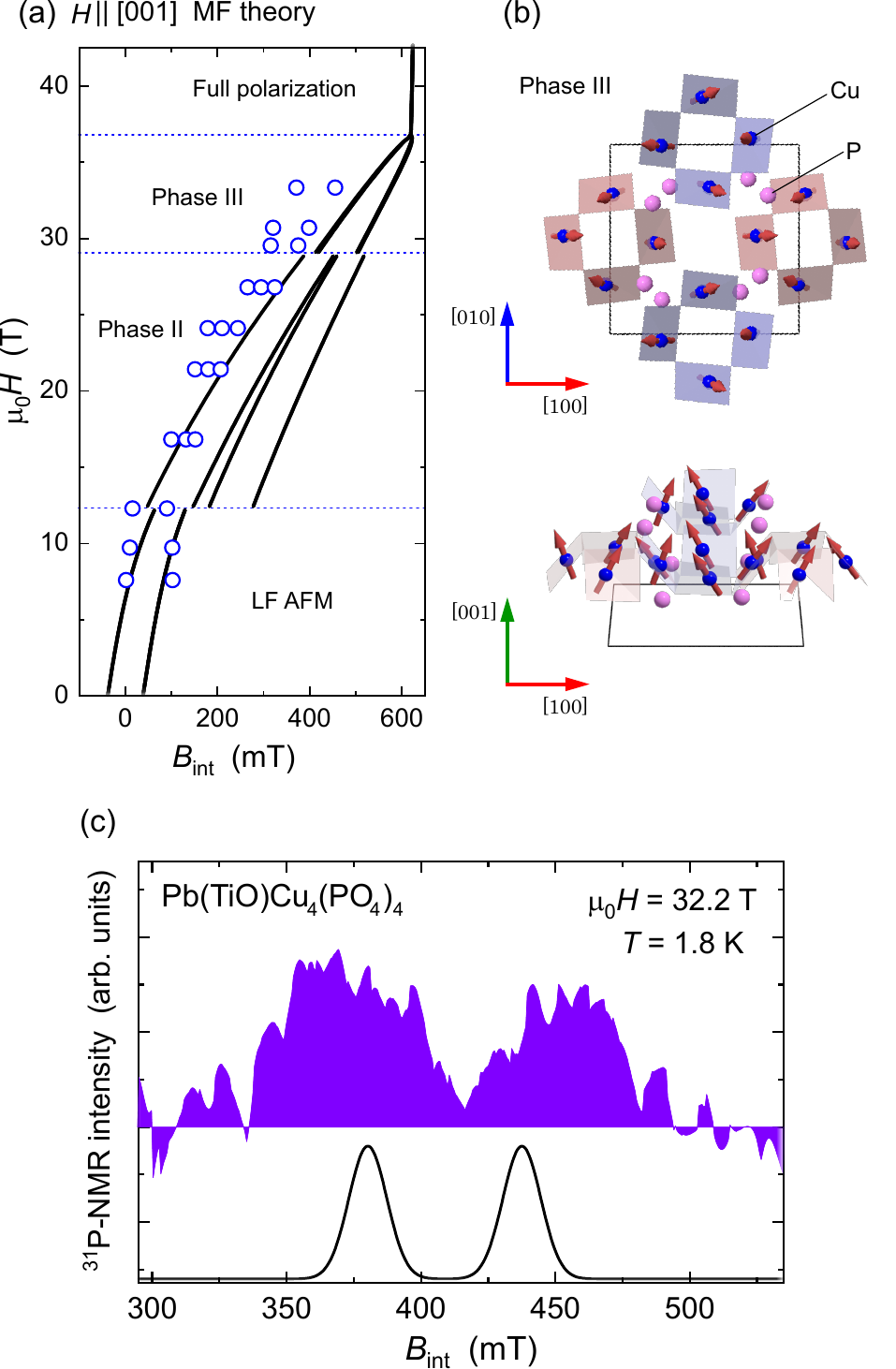}
\end{center}
\caption{(a) Field dependence of $B_{\rm int}$ simulated on the basis of magnetic structure obtained from classical mean-field theory.
Experimental data are plotted again for comparison.
Two $B_{\rm int}$ in phase III \cite{Kato-PRL2017} appearing above 29 T explain the double-peak NMR spectrum.   
(b) Magnetic structure of phase III projected on different planes. 
(c) Comparison of experimental and simulated NMR spectra.
For the simulated spectrum, a linewidth of 10 mT was introduced and whole spectrum was uniformly shifted by 25 \% to compensate the anisotropy of hyperfine coupling constants.
} 
\label{pIII}
\end{figure}
The magnetic structures in high magnetic fields were proposed in the framework of cluster mean-field (CMF) theory \cite{Kato-PRL2017}, 
which consistently explains the experimentally measured magnetic and electric properties of $A$(TiO)Cu$_4$(PO$_4$)$_4$.
The internal field at P sites under $H\parallel [100]$ is, thus, calculated based on the magnetic structures suggested by the CMF theory.
The parameter set used for the CMF theory is the same as those in Ref.~\cite{Kimura-PRM2018}, except for a finite rotation angle $\phi=5^{\circ}$, which is introduced to realistically capture the chiral structure.  
For the simulation, the hyperfine fields from the neighboring four Cu$^{2+}$ moments are estimated using the bond dependent $A_i$ estimated above. 
The direct dipole fields from the Cu$^{2+}$ moment within 100 \AA~ from the target P site are calculated from the structure and included in $B_{\rm int}$.  
The theoretically calculated field dependence of $B_{\rm int}$ is shown in Fig.~\ref{Bint_cal}(a) as black lines. 
The field strength used in CMF theory is scaled at the metamagnetic critical field $B_a$ \cite{Kimura-JPSJ2019}, as represented by dashed horizontal line in Fig.~\ref{Bint_cal}(a).  
The four different $B_{\rm int}$ in the LF phase become two $B_{\rm int}$ in the phase Y \cite{Kato-PRL2017} above $B_a$. 
This result is consistent with the double-peak NMR spectra observed above $B_{a}$.
The peak positions found in Fig.~\ref{NMRSP_HF}(a) is plotted in Fig.~\ref{Bint_cal}(a) by red circles after converting $\Delta f$ to $B_{\rm int}$ using the gyromagnetic ratio $\gamma$ for P nuclear spin. 
Although qualitative field variation is consistently explained, the peak separation is slightly underestimated in the model calculation.    
This is attributed to the offdiagonal component of the hyperfine coupling for the phase Y, which is compensated in the LF phase with higher symmetry. 
The diagonal components do not contribute to the peak separation, but rather shift all peaks in parallel. 
The qualitative agreement between theoretical and experimental results validates the present approach to reveal the magnetic structures in high fields.

Now, $B_{\rm int}$ for $H\parallel[001]$ is calculated following the same scheme used for $H\parallel[100]$ and the result is shown in Fig.~\ref{Bint_cal}(b) as black lines. 
The peak positions found in the NMR measurement are plotted together by blue circles. 
The double-peak spectrum experimentally observed in LF phase is consistently explained by two kinds of $B_{\rm int}$ obtained below $B_{c1}$. 
Four $B_{\rm int}$ appear above $B_{c1}$ associated with the change in the magnetic structure to the phase II \cite{Kato-PRL2017}. 
Experimentally, a central peak and broad tails at both sides were observed at the corresponding field range. 
For these spectra, the peak positions were obtained by fitting the spectra with a three-peak Gaussian. 
We interpret that a peak at the center stems from overlapped two $^{31}$P sites with similar values of $B_{\rm int}$ and the broad tails are assigned to the other two $^{31}$P sites affected by larger and smaller $B_{\rm int}$. 
The model calculation overestimated the peak separation in this case due to the offdiagonal components of hyperfine coupling. 
It should be noted that the double-peak structure appearing above 29 T qualitatively contradicts with the model calculation assuming the phase II. 

To explain the double-peak spectra observed above $B_{c2}$, 
we carried out calculations using the conventional mean-field (MF) theory, in which individual spins are decoupled. 
Since the intercupola interactions in \PbC\, is expected to be stronger than that in \BaC \cite{Kimura-PRM2018,Kato-PRL2017}, the MF theory could be more appropriate for this analysis than the CMF theory. 
The model used here is essentially the same as in the previous study \cite{Kimura-PRM2018}, except that the interlayer coupling is neglected. 
The model parameters were slightly tuned to better reproduce the magnetization curve with the MF calculation \cite{Kato-param}. 
This analysis indicates that phase III \cite{Kato-PRL2017,Kato-PRB2019,Nomura-PRB2023} is stabilized above $B_{c2}$.
The field variation of $B_{\rm int}$ was calculated on the basis of the obtained magnetic structures and represented in Fig.~\ref{pIII}(a). 
The metamagnetic transition to phase III \cite{Kato-PRL2017} appears around 29 T, above which the double-peak structure in NMR spectrum with two $B_{\rm int}$ is suggested. 
The calculated magnetic structure in phase III is illustrated in Fig.~\ref{pIII}(b). 
The field dependence of $B_{\rm int}$ qualitatively explains the experimental result but the average values of $B_{\rm int}$ was overestimated by approximately 25 \%. 
This can be attributed to the angle dependence of hyperfine coupling constant, as $A_p$ for [001] direction is 12 \% smaller than that for isotropic term. 
After correcting $B_{\rm int}$, an NMR spectrum simulated for phase III is compared with the experimental spectrum at 32.2 T in Fig.~\ref{pIII}(c). 
For the simulated spectrum, two peaks at large and small $B_{\rm int}$ are overlapped within a finite linewidth of 10 mT.
The resulting double-peak structure is consistent with the experimental spectrum, which invokes a possibility that the phase III is actually stabilized above 29 T in \PbC. 

Finally, we mention an alternative scenario to stabilize the phase III in high fields. 
In the CMF theory, the phase III can be stabilized at a finite temperature near $T_N$ by reducing the chiral rotation angle of cupola clusters to $\phi=0^\circ$ [Fig.~\ref{structure}(a)] \cite{Kato-PRL2017}, although it never appears in the phase diagram for a realistic value of $\phi \simeq 5^\circ$.
As the phase III becomes more stable by reducing $\phi$, we speculate that the chiral rotation of cupola cluster is suppressed in high magnetic fields due to the magnetostriction effect.  
To directly probe the local magnetostriction effect, the neutron and x-ray diffraction studies in high magnetic fields should be carried out.

\section{Conclusion} \label{con}

We measured the $^{31}$P-NMR spectra for  a single crystalline sample of \PbC\, and estimated the microscopic parameters that describe the coupling between $^{31}$P nuclear moments and Cu$^{2+}$ magnetic moments. 
With these coupling constants the magnetic structure in LF phase was investigated from the NMR spectrum observed below $T_N \simeq 7$ K. 
Furthermore, the $^{31}$P-NMR spectrum measurement was performed at higher magnetic fields reaching 32.2 T using the pulsed high-field technique. 
The metamagnetic transitions and associated modification of magnetic structure were detected as the variation of the NMR spectral shape. 
The internal field at the P site is simulated as a function of external field based on the CMF theory and the result is compared with the obtained NMR spectra. 
The field dependence of peak positions is consistently explained except for the fields higher than $B_{c2}=29$ T applied along [001]. 
We propose that the spin reorientation occurs while entering the phase III, which leads to a double-peak NMR spectrum as observed in experiment and simulated by classical mean-field theory.

\begin{acknowledgements}
This work was carried out under the Visiting Researcher's Program of the Institute for Solid State Physics, the University of Tokyo (ISSPkyodo 202012-HMCXX-0006, 202112-HMCXX-0025).
T.K. was supported by a JSPS Research Fellowship. 
This work was partially supported by JSPS KAKENHI (Grant Nos. 19H01832, 21H01035, 22K03509, 23KJ0502, 24K00575, 24H01599, and 25H00600). 
\end{acknowledgements}


\begin{thebibliography}{99}
\bibitem{Kato-param}
Tha parameters used for the MF theory are $D=0.8, \theta = 100^{\circ},$ and $\phi=2^{\circ}$ in the Hamiltonian identical to that for the CMF theory \cite{Kato-PRB2019}.
\end{thebibliography}
\end{document}